\documentclass[showpacs, preprintnumbers, amsmath, amssymb,twocolumn]{revtex4}

\usepackage{graphicx}
\usepackage{dcolumn}
\usepackage{bm}

\begin{document}

\preprint{...}

\title{The Hall effect of dipole chain in one dimensional Bose-Einstein condensation}

\author{Tieyan Si}\email{sity@itp.ac.cn}\author{Yue Yu}
\affiliation{Institute of Theoretical
Physics, The Chinese Academy of Sciences, Beijing, 100080, China\\
Graduate School of the Chinese Academy of Sciences, Beijing,
100039, China}

\date{\today}

\begin{abstract}

We find a breather behavior of the dipole chain, this breather
excitation obey fractional statistics, it could be an experimental
quantity to detect anyon. A Hall effect of magnetic monopole in a
dipole chain of ultracold molecules is also presented, we show
that this Hall effect can induce the flip of magnetic dipole
chain.

\end{abstract}


\maketitle

\subsection{Introduction}

Fractional Hall effect occurs for electrons confined to two
dimensions in a strong magnetic field\cite{prange}. The Laughlin
wave function has provided a correct description to this quantum
Hall liquid\cite{laughlin}, and it leads us to new profound
understanding to quantum many body system.

The rapidly developed optical method in controlling ultracold cold
molecules gas has opened an unique door to the quantum many body
system\cite{ketterle}. There has been some proposals to realize
the strongly correlated states of fractional Hall type in optical
lattice, such as rotating cold atom gases\cite{wilkin}, melting a
Mott-insulator state in a superlattice potential\cite{anders}, and
so on. While in our paper, a totaly different physical system to
the realization of quantum Hall state in molecule is proposed
based on the dipolar interaction between ultracold molecules.

The Bose-Einstein condensation(BEC) with dipole-dipole
interactions has been thoroughly studied\cite{yiyou1} since the
dipole-dipole interaction was introduced into ultracold molecular
gas by Yi and You\cite{yiyou2}. Yu shows two column of parallel
dipole molecules has an inverse square interaction\cite{yu}, which
happened to be described by the well known Calogero-Sutherland
Model\cite{ca}, the solution of this model is just the Laughlin
wave function. So far, the quantum Hall states of ultracold
molecular clouds with dipole-dipole interaction is still blank to
our acknowledge. So we shall show how the Hall effect arise from
the ultracold molecular dipole chain, the anyons may also be
observed in our physical system.

In this paper, we find a method of controlling the magnetic dipole
momentum through quantum Hall effect. Hall effect can induce the
flip of magnetic monopole in two column of ultracold molecular
dipole chain. The anyon appears as the breathers mode of the
dipole chain, they obey fractional statistics.

\subsection{Hall effect of electric dipole chain}

The BEC may be achieved in a dilute gas of ultracold polar
molecules\cite{molecule}. When the deBroglie wavelength becomes
comparable to the inter-molecule spacing, molecules lose their
individuality, forming a Bose-Einstein condensate. In this case, a
collection of millions of molecules can then be described as a
single entity: a coherent field\cite{hau}. Applying the
'cloverleaf' magnetic trap, which has a weak confinement in $y$
dimension, but very strong along the other two dimensions, the
molecule cloud may be squeezed into a cigar-shaped with an aspect
ratio of about 15:1\cite{mewes}.

At zero temperature, when the confining field becomes stronger,
the cigar shaped molecule cloud would grow thinner, until the
molecules sit in a line one by one along the $y$ direction.
Considering one column of ultracold molecules placed on the line
$x=x_{2},y\in(-\infty,+\infty),z=0$, they are polarized by a
strong homogeneous electric field $E$ penetrating through the
$x-y$ plane with an angle $\theta_{0}$ to the $+x$-axes(Fig. 1).
So there is a electric dipole chain at $x_{2}$, they stand
shoulder to shoulder in a column along $y$ axes. The nearest
neighbor dipole-dipole interaction is repulsive, this keeps the
stability of the cigar-like BEC.

Besides the homogeneous electric field penetrating through the
$x-y$ plane, we also set up a homogeneous magnetic field in the
$y$ direction, it goes perpendicularly into the $x-z$ plane.

\begin{figure}
\begin{center}\label{shall}
\includegraphics[width=0.5\textwidth]{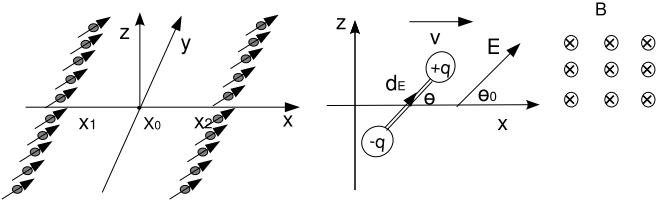}
\caption{Two column of electric dipoles at $x_{1}$ and $x_{2}$
stands shoulder to shoulder along the $y$ axes, each one may be
viewed as a giant dipole. The magnetic field goes perpendicularly
into the paper. As the dipole chain at $x_{2}$ moves to $+x$
direction, the Lorentz force will strengthen the electric dipole,
since they stand shoulder to shoulder, the repulsive force between
them increases, then the dipole chain grows longer.}
\end{center}
\end{figure}

We consider two column of dipole chains at $x_{1}$ and
$x_{2}$(Fig. 1). We first study the electric dipole chain at
$x_{2}$ to investigate macroscopic behavior of the dipole chain.

The electric dipole moment is defined from the electric
polarization charge ${d}_{E}=q\textbf{d}$, $\textbf{d}$ is the
vector pointing from $-q$ to $+q$ polarization electric charge.
When the electric polarization charge $+q$ is moving in the pane
perpendicular to the magnetic field, it feels a Lorentz force,
\begin{equation}\label{lorentz00}
\vec{f}_{e}=q\vec{v}\times\vec{B},
\end{equation}
$\vec{v}$ is the speed of the charge. For the electric dipole
chain at $x_{2}$, its speed is $v_{2}=\frac{dx_{2}}{dt}$. The
Lorentz force's direction obey the right hand rule. When
$v_{2}>0$, according to Eq. (\ref{lorentz00}), the $+q$ would move
towards the $+z$-direction. While the corresponding negative
charges $-q$ of the dipole are moving toward the $-z$-direction.
In the ultracold molecule, the electron cloud moves towards to
$-z$-direction. This strengthened the electric dipole momentum,
therefore the repulsive interaction between the nearest
neighboring electric dipole would increase, then the cigar-like
giant particle grows longer. On the contrary, when the dipole
chain moves in the $-x$-direction, the dipole chain grows shorter.

The direction of the moving electric dipole is given by the
balance between the lorentz force of magnetic field and electric
field. Let $\theta_{0}$ denote the angle between the electric
field and $x$-axes, and $\theta$ denotes the direction of the
electric dipole momentum, then
\begin{equation}\label{tantheta}
\tan{\theta}=E\tan{\theta_{0}}+\frac{vB}{E\cos{\theta_{0}}}.
\end{equation}
This provide us a new method to control the direction of the
electric dipole chain. One need to adjust the ratio between
electric field and magnetic field, and the velocity of the dipole
chain.

\subsection{Breather modes of Calogero-Sutherland gas}

We consider two dipole chains at $x_{1}$ and $x_{2}$(Fig. 1), they
are oscillating around the equilibrium point. Then the dipole
would grows longer and shorter periodically, it will be shown that
this breather mode open a new door to anyon.

The interaction between two parallel electric dipoles is
\begin{equation}\label{Epole}
V_d({\bf y},{\bf
y}')=\frac{E^{2}}{4\pi\epsilon_{0}}\frac{1-3\cos^2\alpha}{r^3},
\end{equation}
with ${\bf y}$ and ${\bf y}'$ as their location and $\alpha$ is
the angle between ${\bf r}={\bf y}-{\bf y}'$ and ${\bf d}$. If the
harmonic trap along the $y$-direction is very flat, $\rho_0(y)$
may be approximated by the average density $\bar\rho_0$. When the
angle between the electric dipole and the $x-y$ plane is $\theta$,
the interaction potential between the two column of dipole chain
is
\begin{eqnarray}
V=\frac{2d^2\bar\rho_0N_0}{2}\biggr(1+\frac{3\pi\cos{\theta}}{4}\biggr)\frac{1}{|x_1-x_2|^2}\label{3},
\end{eqnarray}
where $N_0$ is the molecule number in a condensate, $d$ is the
magnitude of the electric dipole moment ${\bf d}$. In the
following, we denote
$g=2d^2\bar\rho_0(1+\frac{3\pi\cos{\theta}}{4})$ as the coupling
factor. This interaction between the two dipole chain can be
easily controlled through Eq. (\ref{tantheta}).

At zero temperature, the weakly interacting ultracold molecules of
the dipole form one dimensional BEC, they act like a whole to
external perturbation. So each dipole chain may be treated as one
giant particle, whose position on the $x$-axes is marked by
$x_{i}$. The Hamiltonian of this physical system is
\begin{eqnarray}\label{H-cs}
H_{CS}&=&\frac{1}{2}\sum_{i=1}^{2}\biggl(-\frac{d^2}{dx^2_i}
+\omega^2x_i^2\biggr)+\frac{1}{2}\frac{\lambda(\lambda-1)}{|x_1-x_2|^2},
\end{eqnarray}
here we have set the units $m=\hbar=c=1$,
$\lambda=\frac{1-\sqrt{1-4g}}{2}$. Using the operator
decomposition method in supersymmetry quantum mechanics, we find
the operator of the breather excitation,
\begin{equation}\label{hi=d+A}
h_j=-i\frac{d}{dx_j}+A,\;\;\;A=-i{\omega}x_j+i\lambda\frac{1}{x_j-x_k},\;\;
(j\neq{k}),
\end{equation}
$A=(A_{1},\;A_{2})$ is the vector potential. So it is easy to see
the Calogero-Sutherland model for two giant particles describes
the same physics as a two dimensional electron gas. The
Calogero-Sutherland model (\ref{H-cs}) reduced to\cite{ca,suth},
\begin{equation}\label{CS=hh}
H_{CS}=\frac{1}{2}\sum_j h^\dag_j h_j+\hbar\omega(\lambda+1),
\end{equation}
with the ground state energy $E_g=\hbar\omega(\lambda+1).$ Here
$h_j^{\dag}$ is the generating operator of a kind elementary
excitation---the breather excitation. It has an exact ground state
wave function obtained from $h_j|\Psi_{0,\lambda}\rangle=0$,
\begin{eqnarray}\label{psi0}
\Psi_{0,\lambda}(x_1,x_2)=(x_1-x_2)^{\lambda}e^{-\frac{1}{2}\omega^2(x_1^2+x_2^2)}.
\end{eqnarray}
The statistics of the two dipole chain giant particle relies on
$\lambda$. For $\lambda=2k$, the wave function is symmetric when
the two giant particles exchange their position, then they behaves
like bosons. When $\lambda=2k+1$, the wave function is
anti-symmetric for the exchange of the two particles. More over,
$\lambda$ is fractional number, the two giant particles obey
fractional statistics.

The breather excitation mode is also modified by the statistics of
the two giant particles. According to the ground state energy, Eq.
(\ref{CS=hh}), the normal frequency $\omega$ of the breather is
modified by the interaction between the two giant particles,
\begin{equation}\label{w'}
\omega'=\omega(\lambda+1),
\;\;\;\lambda=\frac{1-\sqrt{1-8d^2\bar\rho_0(1+\frac{3\pi\cos{\theta}}{4})}}{2}.
\end{equation}
This excitation spectrum can be probed by measuring the
frequencies of collective oscillation in
experiment\cite{collective}.

\begin{figure}
\begin{center}\label{monopole}
\includegraphics[width=0.35\textwidth]{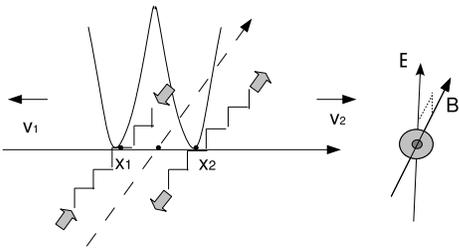}
\caption{When the dipole moves towards the $+x$ direction, the
dipole chain grows longer, when its velocity is in the opposite
direction, it shrinks and becomes shorter. When it oscillates
around the equilibrium point, it behaves like breather, and grows
longer and shorter periodically.}
\end{center}
\end{figure}
If $g<\frac{1}{4}$, the ground state is well defined, according to
Eq. (\ref{w'}), the critical angle of the dipole reads
\begin{equation}\label{theta}
\cos{\theta}<\frac{4}{3\pi}(\frac{1}{8d^{2}\bar{\rho_0}}-1).
\end{equation}
This puts a constrain on the electric and magnetic field. As we
know, if $\lambda$ continuously changed from an odd number to an
even number, another kind of excitation---anyon---would
appear\cite{wilczek}, so this is a good system to detect anyons,
the electric field, magnetic field are both macroscopic physical
parameter. For a fixed value of $\lambda$, combing Eq.
(\ref{tantheta}) and Eq. (\ref{theta}), we can find the proper
range of magnetic field and electric field. In experiment, it is
more convenient to move the electric field in the horizontal plane
than to move the dipole chain.

The above discussion only concerns about two giant particles, in
fact, it also holds for the more general case of $N$ giant
particles. For $n$ giant particles, the physical system is
governed by the Calogero-Sutherland model
$H_{CS}=\frac{1}{2}\sum_i(-\frac{d^2}{dx^2_i}
+\omega^2x_i^2)+\frac{1}2\sum_{i\ne
j}\frac{\lambda(\lambda-1)}{|x_i-x_j|^2}.$\cite{yu}, its ground
state wave function is of a familiar form of Laughlin wave
function. So there are also breathers for the many dipole chain
system, they obey fractional statistics.

\subsection{Hall effect of magnetic dipole chain}

In fact, in our physical system, the magnetic field goes
perpendicular to the $x-z$ plane, so the magnetic dipole chain
lies head to tail in the $y$ direction. But it has been proved
that, for two dipole chain polarized in this way,  the interaction
between the two is zero\cite{yu}, no matter it is static or in
motion.

As analyzed in the above, when the electric dipole chain is
oscillating around the equilibrium point, it induced a breather
mode. If the ultracold molecules has magnetic dipoole momentum,
this dipole chain is a combination of electric dipole pole
chain(like centipede along $y$ axes) and magnetic dipole
chain(head to tail along $y$ axes). While the electric dipole
momentum is much larger than the magnetic dipole momentum. So the
magnetic momentum interaction in one dipole contributes a small
perturbation. Though the magnetic dipole has a small attractive
interaction, which is opposite to the repulsive interaction
between electric dipoles, it also present the same breather
behavior as the electric case.

We shall introduce Dirac's magnetic monopole\cite{dirac} to study
this self-consistence. A magnetic dipole may be viewed as a
monopole-anti-monopole pair, the magnetic momentum is defined by
the product of the monopole charge and vector from the
anti-monopole to the monopole, ${\mu}_{m}=g\textbf{d}$, $g$ is the
charge of magnetic monopole, $\textbf{d}$is the vector pointing
from anti-magnetic-monopole $-g$ to positive magnetic monopole
$+g$.

The Maxwell equations for a vacuum without sources possess an
interesting symmetry under the exchange of electric field E and
magnetic field B, i.e., $E\rightarrow{B}$ and $B\rightarrow{-E}$,
this symmetry is called the electric-magnetic duality. So for a
magnetic monopole with charge $g$ moving in the plane
perpendicular to electric field, it feels a dual Lorentz force,
\begin{equation}\label{lorentz01}
\vec{f}_{g}=-g\vec{v}\times\vec{E},
\end{equation}
this duality is self-content during the movement of the dipole
chain. The dipole chain is a series of monopole-anti-monopole pair
head to tail along $y$ direction.  As the magnetic dipole chain
moves in $+x$ direction, the monopole $+g$ in the electric field
is govern by the dual Lorentz force (\ref{lorentz01}), so there is
a transverse force towards $+y$ direction acted on the magnetic
monopole, the anti-monopole $-g$ would move to $-y$-direction
correspondingly. This strengthen the cancellation of the monopole
and anti-monopole charge in the middle of dipole chain except the
charges at the two end points of the giant particle. Monopoles
with $+g$ are focused on the $+y$ end point, and anti-monopole
with $-g$ are focused on the $-y$ end point. This is the Hall
effect of magnetic monopoles.

There is also a corresponding Hall current of magnetic monopole,
the corresponding Hall conductance is determined by the net force
include the Lorentz force
\begin{equation}\label{lorentz}
\vec{f}_{g}=g(\vec{B}-\vec{v}\times\vec{E}),
\end{equation}
$\vec{v}$ is the charge velocity. The balance $f=0$ provides a
relation between the magnetic field along the y-direction $E_{y}$
and the applied magnetic field $E_{z}$,
\begin{equation}\label{E=vB}
B=v^{0}_{x}E_{z}.
\end{equation}
When the velocity of the dipole chain exceed the critical velocity
$v>v^{0}$, the net force would change its sign, in that case, the
magnetic dipole chain would turn to the opposite direction. When
the velocity of the giant particle is twice of the critical
velocity $v^{c}$, there would be an opposite dipole chain with the
same strength. This has provide us a new method a obtain a
antiparallel dipole array. This flip of magnetic dipole momentum
is induced by the its movement in electric field.
\begin{figure}
\begin{center}\label{hall}
\includegraphics[width=0.35\textwidth]{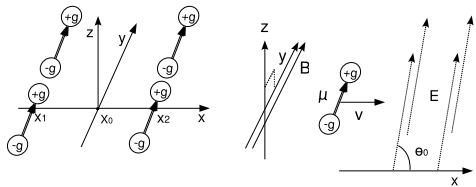}
\caption{The magnetic dipole momentum is pined to the $y$
direction by the magnetic field. When the dipole chain moves
towards to $+x$ direction, the positive magnetic dipole follows
the left hand rule, and was attracted to $+x$ direction, while the
negative monopole is pushed to the $-x$ direction, then the dipole
chain grows longer.}
\end{center}
\end{figure}

The Dirac quantization condition says that in the presence of
magnetic monopoles, the product of electric and magnetic charges
must be an integral multiple of $1/2$, i.e., $e^{-1}=2g_{0}/n,
(n=1,2,3,...)$. So the dual Hall resistance of magnetic monopole
$R_{g}$ reads
\begin{equation}\label{hallR=lam}
R_{g}=\lambda\frac{h}{n}{4g_{0}^{2}}=\frac{1}{\nu}\frac{h}{n}{4g_{0}^{2}}.
\end{equation}
here $\nu$ is the filling factor of Landau level,
$g_{0}=\frac{1}{e}$ is the unit of magnetic monopole charge, $e$
is the unit electric charge. The magnetic monopole has been
drawing great attention of physicists all the time since Dirac
postulated its existence in 1931\cite{dirac}. There is still no
definite evidence to confirm its existence in spite of the great
effort physicists have made in the last decades of years. This
Hall effect would provided us a new understanding to magnetic
monopole.

In fact, Eq. (\ref{lorentz01}) means the magnetic monopole obey
the left handed rule, it leads to self-content results with
electric dipole chain. On the contrary, if magnetic monopole obey
the right handed rule, when the magnetic dipole is lengthened, the
electric dipole chain would be shortened, this two motions are
conflict. Which case is correct on earth? it lies in the hand of
exact experiment.

\subsection{conclusion}

In summary, when two column of ultracold molecules of
Bose-Einstein condensation are polarized by an orthocline magnetic
field and electric field, they become two dipole chains. When the
dipole is oscillating in the transverse section, it induces a
breather excitation in the longitude direction. The length of
dipole chain behaves like a breather, it grows longer and shorter
periodically. This breather excitation obey the fractional
statistics. This could be an experimental quantity to detect the
anyon. By adjusting the speed of the dipole chain, one can observe
a magnetic dipole flip induced by Hall effect above a critical
velocity.

\subsection{Acknowledgment}

We are grateful to Prof. S. J. Yang,  Dr. Y. C. Wen, Dr. X. C. Lu,
and Dr. J. B. Li for stimulating discussions. This work was
supported by the National Natural Science Foundation of China.

\end{document}